\documentclass[twocolumn]{revtex4-1}
\usepackage[utf8]{inputenc}
\usepackage{amsmath}
\usepackage{amssymb}
\usepackage{braket}
\usepackage{graphicx}
\usepackage{float}

\begin{document}

\title{Elliptic Curves in Continuous-Variable Quantum Systems}
\date{September 2022}

\author{Maxwell Aifer}
\affiliation{Department of Physics, University of Maryland, Baltimore County, Baltimore, MD 21250, USA}

\author{Evan Sheldon}
\affiliation{Department of Physics, University of Maryland, Baltimore County, Baltimore, MD 21250, USA}

\date{\today}

\begin{abstract}

Elliptic curves are planar curves which can be used to define an abelian group. The efficient computation of discrete logarithms over this group is a longstanding problem relevant to cryptography. It may be possible to efficiently compute these logarithms using a quantum computer, assuming that the group addition operation can be computed efficiently on a quantum device. Currently, however, thousands of logical qubits are required for elliptic curve group addition, putting this application out of reach for near-term quantum hardware. Here we give an algorithm for computing elliptic curve group addition using a single continuous-variable mode, based on weak measurements of a system with a cubic potential energy. This result could lead to improvements in the efficiency of elliptic curve discrete logarithms using a quantum device.
\end{abstract}

\maketitle

\emph{Introduction.---}Elliptic curves have a prominent role in number theory, and are of great practical importance in modern cryptography, as they provide an alternative to the RSA algorithm \cite{bernstein2017post, washington2008elliptic, koblitz1993introduction}. This makes the efficient computation of the elliptic curve group operation (and the discrete logarithm over this group) a question of great importance. As alternative computing paradigms can in some cases provide speed-ups over classical digital computers, it is of interest to explore the computation of elliptic curve group operations in different paradigms.

Previous work has explored the use of discrete-variable and quantum computers to evaluate the elliptic curve group operation and take discrete logarithms over this group; however, to use these results in practical scenarios requires thousands of logical qubits \cite{haner2020improved}, placing it beyond NISQ-era capabilities. Continuous-variable quantum information is seen as a promising approach to realizing cryptographic protocols like quantum key distribution \cite{jain2022practical, weedbrook2012gaussian}, as continuous-variable states are compatible with existing telecom infrastructure; however, to our knowledge, no attempt has been made to evaluate the elliptic curve group using continuous variable quantum systems. There has also been significant effort to design classical computing hardware for elliptic curve group addition \cite{verri2020review}.

In this work, we propose a method for evaluating the group operation of an elliptic curve (over the reals) using a continuous-variable quantum system. As is well known, the elliptic curve group addition operation can be cast as a geometric relationship between points in the plane, in particular by identifying points of intersection between a straight line and an elliptic curve \cite{washington2008elliptic, koblitz1993introduction}. We show that a continuous-variable quantum system can be designed whose energy eigenstates have Wigner functions resembling elliptic curves, by implementing a Hamiltonian with a cubic potential energy function. Moreover, weak quantum measurements of quadrature operators can effectively project this Wigner function onto a straight line in the plane, resulting in a method for elliptic curve point addition.

Interestingly, this algorithm is entirely based on the geometric properties of a quantum system's Wigner function, and those of the weak measurement operation. We note that existing algorithms of elliptic curve group addition work by operating on binary encodings of elliptic curves, and these encodings do not readily exhibit the geometry of the problem. In contrast, our algorithm operates directly on an object that geometrically embodies the elliptic curve, namely the Wigner function of a continuous-variable mode. We therefore consider this algorithm exemplary of a novel "geometric computing" paradigm, which may be applicable both in classical and quantum settings.

It is also shown that a superconducting nonlinear asymmetric inductive element (SNAIL) device \cite{PhysRevLett.125.160501, frattini20173} can be used to realize the necessary cubic potential physically, suggesting a path to the experimental implementation of our algorithm.

\begin{figure}[t]
    \centering
    \includegraphics[width=8.6cm]{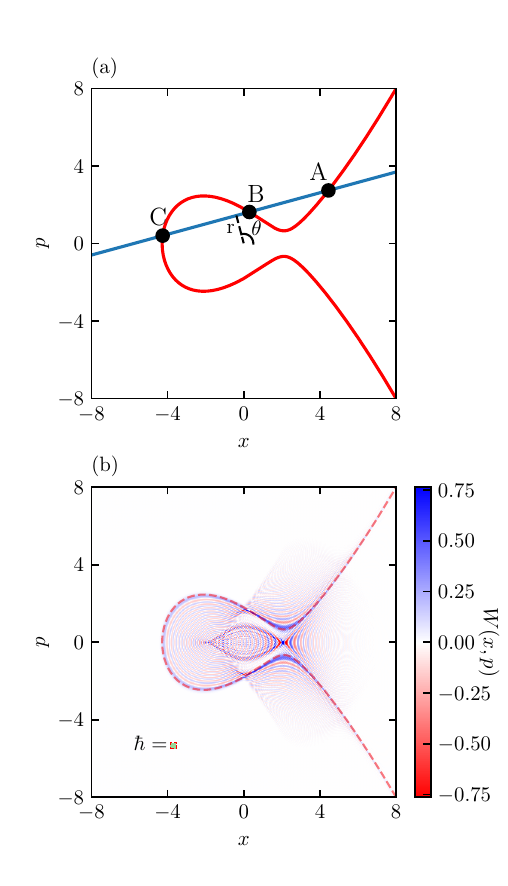}
    \caption{(a) The elliptic curve corresponding to the classical Hamiltonian \eqref{eq:classical-hamiltonian} with parameters $a = 0.075$ and $b=-1$, and $m=1$, and $E=1.62$ (arbitrary units). (b) The Wigner function derived from the Eigenvector of the quantum Hamiltonian \eqref{eq:Schrodinger-Eq} with the same parameters and $\hbar = 0.1$. In some places, there were values of the Wigner function as large as $W=1.53$, but to make the form of the function easily visible, values of the Wigner function are cut off at $\pm 0.75$.} 
    \label{fig:Wigner-function}
\end{figure}

\emph{Elliptic Curves in a Continuous Variable Quantum System.---}A real elliptic curve is the set of points $(x,y)\in \mathbb{R}^2$ defined by the equation \cite{koblitz1993introduction}
    \begin{equation}
    \label{eq:elliptic-curve-def}
        y^{2} = ax^{3} + bx + c,
    \end{equation}
for some $a,b,c \neq 0$. A group can be defined over the points of an elliptic curve. Let $A = (x_A, y_A)$ and $B = (x_B, y_B)$ be two points belonging to an elliptic curve, as in Fig. \ref{fig:Wigner-function}a. We denote the slope of the line connecting $A$ and $B$ by $s = (y_B - y_A)/(x_B - x_A)$. We may compute a third point $C = (x_C, y_C)$ from $A$ and $B$ as follows
\begin{equation}
    \label{eq:elliptic-curve-plus-def-x}
    x_C = s^2 - x_A - x_B,
\end{equation}
\begin{equation}
    \label{eq:elliptic-curve-plus-def-y}
    y_C = - y_A + s(x_A - x_B). 
\end{equation}
Then we may define a binary operation ($+$) acting on points of the curve by
\begin{equation}
    A + B = -C
\end{equation}
where $-C = (x_C, -y_C)$ is the reflection of $C$ over the $x$ axis. It can be shown that the points of the curve form a group under the operation $(+)$. Graphically, we can interpret the group addition rule in the following way: to obtain the point $C$, take points $A$ and $B$, draw a line that intersects the 2 points and also a third point $-C$ on the curve (See Fig. \ref{fig:Wigner-function}a). Finally, $-C$ is reflected over the $x$ axis to obtain $C$. We also define coordinates $r$ and $\theta$ as an alternative way to specify the line between $A$ and $B$. Namely,
\begin{equation}
\label{eq:polar-coords-def}
    \theta = \arctan(s), \: \: \: r = (y_A - s x_A )\cos(\theta)
\end{equation}
Elliptic curves appear naturally as the characteristic phase-space curves of a system with a cubic potential energy. Suppose that a classical system has a potential energy
\begin{equation}
    \label{eq:cubic-potential}
    V(x) = -a x^3 - b x,
\end{equation}
with real constants $a,b >0$. Then the classical Hamiltonian is given by
\begin{equation}
    \label{eq:classical-hamiltonian}
    H(x,p) = -a x^3 - b x + \frac{1}{2m} p^2,
\end{equation}
where $m$ is the mass. Consider the constant energy set obtained by setting $H(x,p) = E$. If we make the identifications $E = c$ and $y = p/\sqrt{2m}$, then Eq. \eqref{eq:classical-hamiltonian} is equivalent to Eq. \eqref{eq:elliptic-curve-def}, so the equal-energy curves are elliptic curves as shown in Fig. \ref{fig:Wigner-function}(a). A corresponding quantum Hamiltonian can be defined by replacing the variables $x$ and $p$ with the dimensionless quadrature operators $\hat{x} = (\hat{a}^\dag + \hat{a})/\sqrt{2}$ and $\hat{p} = i(\hat{a}^\dag - \hat{a})/\sqrt{2}$, where $\hat{a}^\dag$ and $\hat{a}$ are respectively the creation and annihilation operators for a single bosonic mode \cite{serafini2017quantum, schleich2011quantum}. That is, we define
\begin{equation}
    \label{eq:quantum-hamiltonian}
    \hat{H} = -a \hat{X}^3 - b \hat{X} + \frac{1}{2m} \hat{P}^2.
\end{equation}
A Hamiltonian of this form can be realized physically using a SNAIL device \cite{PhysRevLett.125.160501} with the correct choice of parameters. Note that the SNAIL device can interpolate between quadratic and cubic Hamiltonians; therefore one can begin with an eigenstate of the harmonic oscillator Hamiltonian, and then adiabatically deform the Hamiltonian into the the form of Eq. \eqref{eq:quantum-hamiltonian}, which will result in an energy eigenstate of the latter \cite{kato1950adiabatic}.
We now consider an eigenstate of the elliptic curve Hamiltonian, that is a wavefunction $\ket{\Psi}$ satisfying the time-independent Schrödinger equation
\begin{equation}
    \label{eq:Schrodinger-Eq}
    \hat{H}\ket{\Psi} = E\ket{\Psi},
\end{equation}
where we have set $\hbar = 1$. In particular, we consider the Wigner function of such an energy eigenstate, defined as \cite{schleich2011quantum}
\begin{equation}
\label{eq:wigner-function-def}
    W(x,p) = \int_{-\infty}^{\infty}dy e^{ipy}\Braket{x-\frac{y}{2}|\Psi} \Braket{\Psi|x + \frac{y}{2}}.
\end{equation}
The Wigner function is a quasiprobability distribution which shares certain features with classical phase space probability density function, notably that integrating it in any direction results in a valid marginal probability density function. We therefore intuitively expect that the Wigner function $W(x,p)$ of the energy eigenstate $\ket{\Psi}$ will be concentrated around the classical phase-space orbit corresponding to energy $E$. In Fig. \ref{fig:Wigner-function}, we see that this is in fact the case; we have plotted both the classical phase space orbit in panel (a), and the Wigner function of the quantum Hamiltonian in panel (b), and remark that the Wigner function of the classical system shows high concentration around the classical phase space orbit. There are also regions where the Wigner function has support that are distant from the classical phase space orbit. However, one notices that in these regions the Wigner function oscillates rapidly, meaning that they will often be eliminated from the marginal distributions due to destructive interference.
The Wigner function was obtained in two steps. First, the eigenstate $\ket{\Psi}$ was constructed in the position basis by discretizing the Hamiltonian operator and then numerically performing an eigendecomposition. Then the Wigner function was evaluated on a grid via straightforward numerical integration of Eq. \eqref{eq:wigner-function-def} using methods similar to those described in \cite{Stack_Exchange}. We also adapted code from \cite{Stack_Exchange} for generating plots of the Wigner function.

Having found a way to realize quantum states that are representative of elliptic curves, we next will turn to the problem of approximately evaluating the elliptic curve group addition operation using this state.

\emph{Evaluation of the Elliptic Curve Group Operation.---}The problem we would like to solve is as follows: given two points $A = (x_A, y_A)$ and $B = (x_B, y_B)$, evaluate the coordinates of the point $C = A + B$, where addition is understood as defined in equations \eqref{eq:elliptic-curve-plus-def-x} and \eqref{eq:elliptic-curve-plus-def-y}. Graphically, the addition operation can be carried out by drawing a line through the points $A$ and $B$, then finding the third point where this line intersects the elliptic curve, and finally reflecting the point thus obtained over the $x$ axis. After a measurement of a quadrature operator (that is, a linear combination of the $\hat{X}$ and $\hat{P}$ operators), the Wigner function will ``collapse" so that its support is a line in the phase space. However, performing such a measurement on the elliptic curve state described earlier would leave none of the structure of that state, making it impossible to extract the result of the group addition operation.
Alternatively, we can consider weak measurements of a quadrature operator on an elliptic curve state \cite{nielsen2010quantum, wiseman2009quantum}. Such a weak measurement can be seen as a kind of interpolation between the identity operation (leaving the elliptic curve state unchanged) and a projective quadrature measurement (completely collapsing the wavefunction).
This can be accomplished using weak measurements of a quantum system with the cubic potential energy in Eq. \eqref{eq:cubic-potential}. Specifically, define a basis of quadrature states $\ket{\psi_\theta}$ as the eigenvectors of the quadrature operator $x_\theta$
\begin{equation}
    \hat{X}_\theta \ket{x_\theta} = x_\theta \ket{x_\theta} ,
\end{equation}
where
\begin{equation}
    \hat{X}_\theta = \cos(\theta)\hat{X} + \sin(\theta) \hat{P}.
\end{equation}
To define a weak measurement of quadrature $\hat{X}_\theta$ having outcome $\mu$ and width parameter $w$, we define the Krauss operator
\begin{equation}
\label{weak-measurement-krauss-def}
    K_{\theta} =\frac{2^{1/4}}{( \pi w^2)^{1/4}} \int_{-\infty}^{\infty} dx_{\theta} e^{-(x_{\theta}-\mu)^{2}/w^{2}}\ket{x_{\theta}}\bra{x_{\theta}}
\end{equation}
For a system initialized in the state $\ket{\psi_0}$, the Krauss operator can be used to determine the post-measurement state $\ket{\psi_\text{pm}}$ as follows
\begin{equation}
    \ket{\psi_\text{pm}} = \frac{K_\theta \ket{\psi_0}}{\sqrt{\braket{\psi_0|K_\theta^\dag K_\theta |\psi_0}}}.
\end{equation}
\begin{figure}[H]
    \centering
    \includegraphics[width=8.6cm]{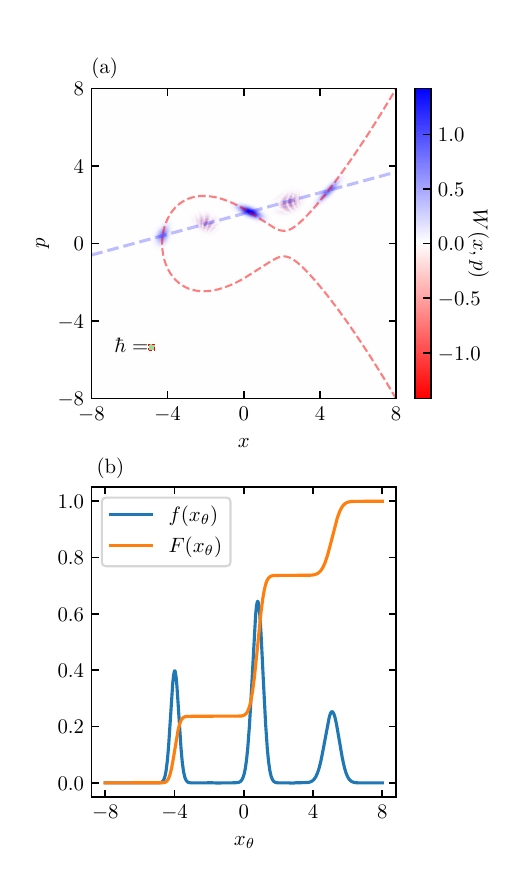}
    \caption{(a) The post-measurement state after a weak measurement of a quadrature operator $\hat{X}_\theta$, with $\theta = \pi/12$ and $r = 1.5$. Similarly to Fig.\ref{fig:Wigner-function}, the range of values for the Wigner function has been restricted, in this case to the interval $\pm 1.42$. In some places, there were values of the Wigner function as large as $W=2.84$.
    (b) Marginal distribution for the second quadrature measurement (of the operator $\hat{X}_{\theta+\pi/2}$). The blue curve $f(x)$ is the probability density function for the marginal distribution, and the orange curve $F(x)$ is the cumulative distribution function, where the three peaks correspond to the three blue regions of concentration in phase space of the Wigner function in Fig. 2a. In Fig. 2a we also see two regions of concentration with wavelike patterns in between these, which are not present in the marginal due to destructive interference.}
    \label{fig:weak-fig}
\end{figure}
\begin{figure}[H]
    \centering
    \includegraphics[width=8.4cm]{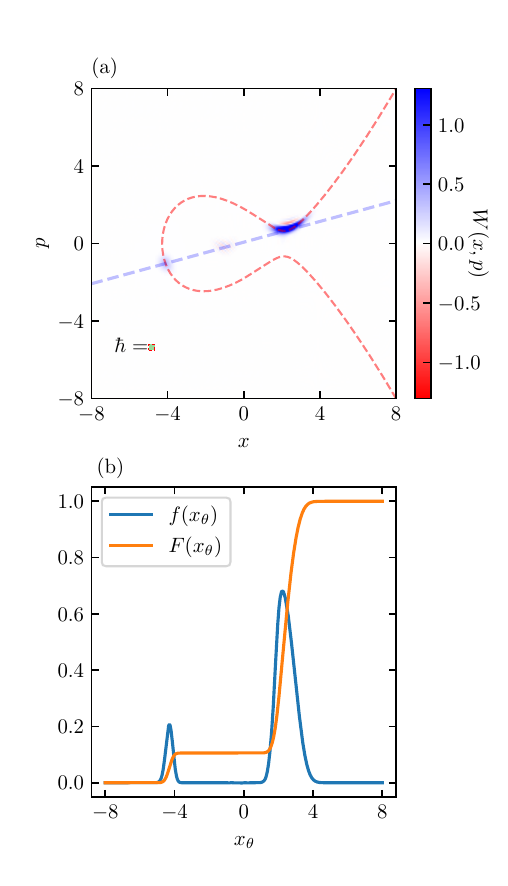}
    \caption{(a) The post-measurement state after a weak measurement of a quadrature operator $\hat{X}_\theta$, with $\theta = \pi/12$ and $r = 0.07$. Similarly to Fig.\ref{fig:Wigner-function}, the range of values for the Wigner function has been restricted, in this case to the interval $\pm 1.31$. In some places, there were values of the Wigner function as large as $W=2.61$. 
    (b) Marginal distribution for the second quadrature measurement (of the operator $\hat{X}_{\theta+\pi/2}$). The blue curve $f(x)$ is the probability density function for the marginal distribution, and the orange curve $F(x)$ is the cumulative distribution function, where the peaks correspond to the blue regions of concentration in phase space of the Wigner function in Fig. 3a.}
    \label{fig:weak-fig2}
\end{figure}

In Figs. \ref{fig:weak-fig} and \ref{fig:weak-fig2} we see the effect of a weak quadrature measurement on the Wigner function of our system, initially prepared in an eigenstate of the Hamiltonian \eqref{eq:quantum-hamiltonian}.

\begin{enumerate}
    \item Prepare an energy eigenstate $\ket{\psi_n}$ of the cubic Hamiltonian.
    \item Compute the polar coordinates $r, \theta$ from Eq. \eqref{eq:polar-coords-def}, and make a weak measurement of the quadrature $x_\theta$.
    \item Choose some tolerance $\delta_r$. If $|x_\theta - r| > \delta_r$, repeat steps 1 and 2 until $|x_\theta - r| \leq \delta_r$. This will require $O(1/\delta_r)$ trials on average.
    \item Take a projective measurement of the quadrature $x_{\theta+\pi/2}$, and compute $x$ and $y$.
    \item Make sure that the computed $x$ and $y$ are not the points $A$ or $B$, and if so, repeat steps 1-4
\end{enumerate}

\emph{Discussion.---}
In summary, we have demonstrated the ability to perform the group addition operation over an elliptic curve group by performing weak measurements on elliptic curve-shaped Wigner functions, which could be realized experimentally using a quantum SNAIL device \cite{PhysRevLett.125.160501}.
Once the validity of our method has been demonstrated through experiments, the algorithm can be modified and extended to accomplish more complex tasks. In particular, it is desirable to extend this method to perform elliptic curve exponentiation \cite{Exponentiation}; if an elliptic curve exponentiation can be performed that preserves a coherent superposition, then Shor's algorithm may be applied to achieve efficient computation of the discrete logarithm, which would have far-reaching impacts in cryptography.

Further investigations into this topic include implementing quantum error-correcting procedures for this continuous variable state. Given that this protocol will be affected by environmental decoherence, the experimental realization of this algorithm would require a protocol for continuous-variable error correction, which has been investigated to some extent \cite{van2010note, dias2018quantum, barnes2004stabilizer}.

It is also important to estimate the resource costs of our quantum algorithm, especially the time and energy needed to evaluate the elliptic curve group addition operation. Analyses of the time and energy costs of discrete-variable quantum computations have been carried out \cite{ikonen2017energy, deffner2021energetic, aifer2022quantum, auffeves2022quantum}, and a similar method could be adapted to continuous-variable systems to address our algorithm.

\bibliographystyle{unsrt}
\bibliography{references}

\onecolumngrid

\appendix

\section{Elliptic Curves}
The material on elliptic curves here is taken from references \cite{koblitz1993introduction} and \cite{washington2008elliptic}.
\label{app:elliptic-curves}
    \begin{figure}[H]
    \centering        \includegraphics[width=\textwidth]{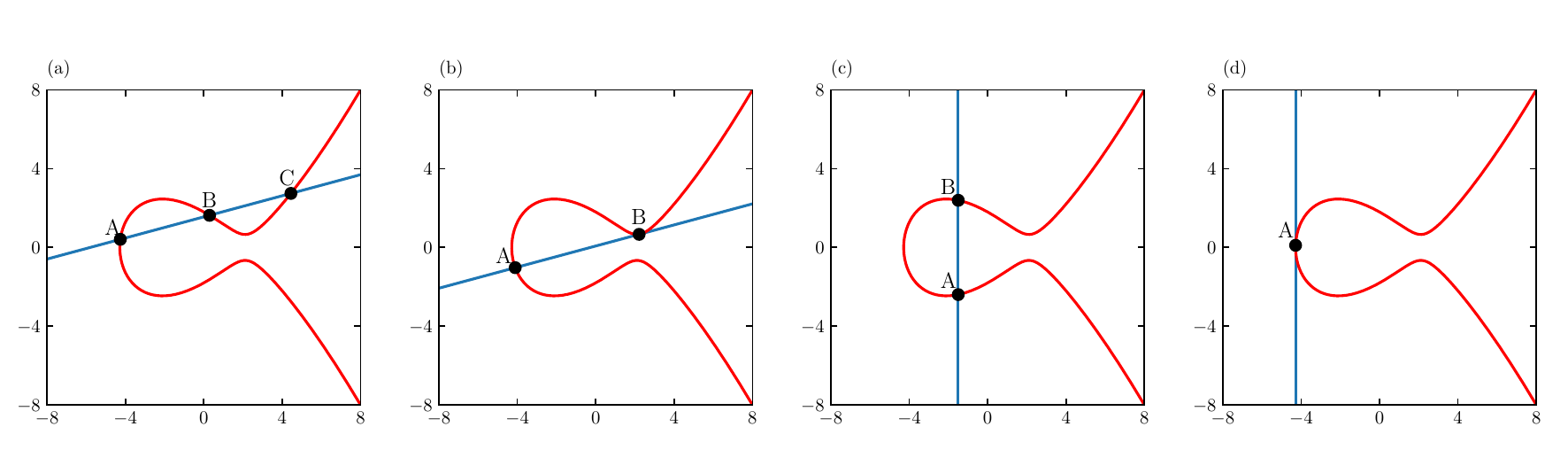}
    \label{fig:elliptic-model-1}
    \end{figure}
    
    An elliptic curve is a curve that is defined by the equation
    \begin{equation}
        y^{2} = ax^{3} + bx + c
    \end{equation}

    If we want to represent the points on an elliptic curve as a group, we can take points A and B, draw a line that intersects the 2 points, and also a third point C on the curve; as displayed in (a). We can then define
    \begin{equation}
        A + B = -C
    \end{equation}
    where \(-C\) is a reflection of \(C\) across the x-axis.

    In the case of (b), the third intersection point C approaches B, and the line becomes tangent to the curve at B.
    As a result, the tangent point, \(B\) is also the intersection point so we can define
    \begin{equation}
        A + B = -B
    \end{equation}

    In the case of \(B = -A\) as seen in (c), we encounter a vertical intersection line through the curve. An issue in this case is that we don't have a third point on the curve our line intersects.
    Our solution is to define an additional point \(\mathcal{O}\) at infinity. An important property is that
    \(\mathcal{O}\) exists on all vertical lines.
    \begin{equation}
        A + B = A + (-A) = \mathcal{O}
    \end{equation}

    The last case (d) we will look over is a vertical line with one point of intersection on the curve, \(B = A\). Since our intersection line is vertical, we can represent this case as
    \begin{equation}
        A + A = \mathcal{O}
    \end{equation}

\section{Phase Space Description}
\label{app:phase-space}
The material on the quantum phase space formalism is taken from references \cite{serafini2017quantum} and \cite{schleich2011quantum}.
    The state of a quantum system is, in general, represented by a density operator $\hat{\rho}$, which may be expanded in an eigenbasis of pure states as
    \begin{equation}
        \hat{\rho} = \sum_{n}p_{n}\ket{\psi_{n}}\bra{\psi_{n}}.
    \end{equation}
    Absent from this description is any concept of the phase space, which is prominent in the corresponding classical theory. However, quantum systems also admit a phase-space description, which can be found using the Weyl transformation. An operator $\hat{A}$ acting on the quantum Hilbert space can be mapped to a phase-space function $\tilde{A}(x,p)$ via
    \begin{equation}
        \tilde{A}(x,p) = \int_{-\infty}^{\infty}dy e^{ipy/\hbar}\Braket{x-\frac{y}{2}|\hat{A}|x+\frac{y}{2}}
    \end{equation}
    So, if we want to be able to map the action of the density matrix in phase space, we should perform a Weyl transformation on the density matrix, and in doing so we can derive the Wigner function.
    \begin{equation}
        W(x,p) = \tilde{\rho}/h = \frac{1}{h}\int_{-\infty}^{\infty}dy e^{ipy/\hbar}\Braket{x-\frac{y}{2}|\hat{\rho}|x+\frac{y}{2}}
    \end{equation}
    One of the key properties of the Wigner function is that we can calculate the probability distribution in space or momentum of the state by integrating along \(p\) or \(x\) respectively.
    \begin{align}
        & \int_{-\infty}^{\infty} W(x,p)dp = \braket{x|\hat{\rho}|x} = \rho(x,x) = P(x) \\
        & \int_{-\infty}^{\infty} W(x,p)dx = \braket{p|\hat{\rho}|p} = P(p) 
    \end{align}
    In fact, the Wigner function can be marginalized over an arbitrary axis in the $x$-$p$ plane to yield a probability distribution for the perpendicular coordinate. That is, if we define the variables $x_\theta$ and $p_\theta$ (called quadratures)
    \begin{equation}
       x_\theta = \cos(\theta)x + \sin(\theta)p, \: \: \: \: p_\theta = -\sin(\theta)x + \cos(\theta)p.
    \end{equation}
    Integrating the Wigner function with respect to one of these quadratures gives the probability density function for the perpendicular one
    \begin{align}
        & \int_{-\infty}^{\infty} W(x_\theta,p_\theta)dp_\theta = \braket{x|\hat{\rho}|x} = P(x_\theta) \\
        & \int_{-\infty}^{\infty} W(x_\theta,p_\theta)dx = P(p_\theta). 
    \end{align}
    Importantly, the Wigner function is not quite a probability distribution, as it can have negative values, and in fact the negativity of the Wigner function characterizes the quantum behavior of a system. For this reason the Wigner function is often referred to as a quasi-probability distribution.

    Although typically a continuous-variable quantum state is expressed in the position or momentum basis, we may also use the basis corresponding to the quadrature variables $x_\theta$ or $p_\theta$. To this end, we define a quadrature operator $\hat{X}_\theta$
    \begin{equation}
        \hat{X}_{\theta} = \cos(\theta)\hat{x} + \sin(\theta)\hat{p},
    \end{equation}
    where we restrict the region of theta to \(0 \leq \theta \leq \pi\). The basis associated with this operator is determined by the eigenvalue equation
    \begin{equation}
    \hat{X}_{\theta}\ket{X_{\theta}} = X_{\theta}\ket{X_{\theta}}.
    \end{equation}
    Multiply both sides by \(\bra{x}\) to achieve
    \begin{align}
        & \braket{x|\hat{X}_{\theta}|X_{\theta}} = \braket{x|X_{\theta}|X_{\theta}} \\
        &\rightarrow \hat{X}_{\theta}\psi(x) = X_{\theta}\psi(x)
    \end{align}
    where \(\psi(x) = \braket{x|X_{\theta}}\).
    \begin{align}
        &\rightarrow (\cos(\theta)\hat{x} + \sin(\theta)\hat{p})\psi(x) = X_{\theta}\psi(x) \\
        &\rightarrow \cos(\theta) x\psi(x) - i\hbar\sin(\theta)\partial_{x}\psi(x) = X_{\theta}\psi(x)
    \end{align}
    The solution to this eigenvalue problem is
    \begin{equation}
        \psi(x)= C_\theta\exp\left[\frac{-ix(-2X_{\theta} + x\cos(\theta))}{2\hbar\sin(\theta)}\right]
    \end{equation}
    Since this wavefunction is imaginary, we need to normalize it in reference to delta functions. By projecting onto the completeness relation and utilizing orthogonality, we yield the normalization constant
    \begin{equation}
        |C_{\theta}|^{2} = \frac{1}{2\pi\hbar\sin(\theta)}
    \end{equation}
    This gives our rotated quadrature eigenstate wavefunction of
    \begin{equation}\label{rotated_wavefunction}
        \psi(x) =\sqrt{\frac{1}{2\pi\hbar\sin(\theta)}}\exp\left[\frac{-ix(-2X_{\theta} + x\cos(\theta))}{2\hbar\sin(\theta)}\right]
    \end{equation}

\section{Weak Measurements}
\label{app:weak-measurements}

    This treatment of weak measurements is based on references \cite{nielsen2010quantum} and \cite{wiseman2009quantum}. If we want to take a weak measurement of a system, we must take the system's initial quantum state \(\ket{\psi}\), and couple it with an ancilla state \(\ket{\varphi}\). 
    This gives us the combined initial system of \(\ket{\Psi} = \ket{\psi}\otimes\ket{\varphi}\).
    The entire system evolves in time through the time evolution operator, represented by \(U(t)\).
    \begin{equation}
        \ket{\Psi_{t}} = U(t)\ket{\Psi_{0}}
    \end{equation}
    
    We can solve for \(U(t)\) by plugging it into the Schrodinger Equation.
    \begin{align}
        & \frac{dU(t)}{dt} = \frac{-i}{\hbar}HU(t) \\
        & U(t) = e^{\frac{-itH}{\hbar}}
    \end{align}
    \(H = H_{S}\otimes H_{A}\) where \(H\) is the Hamiltonian of the whole system, \(H_{S}\) is the Hamiltonian of the quantum state, and \(H_{A}\) is the Hamiltonian of the ancilla state.
    We can also represent \(U(t)\) with a Taylor series expansion. Assume \(t\) is small such that \(t^{3}\approx 0\)
    \begin{align}
        & U(t) = I\otimes I - itH - \frac{1}{2}t^{2}H^{2} + O(t^3) \\
        &\approx I\otimes I - itH_{S}\otimes H_{A} - \frac{1}{2}t^{2}H_{S}^{2}\otimes H_{A}^{2}
    \end{align}
    This allows us to rewrite the time evolved combined system as follows,
    \begin{equation}
        \ket{\Psi_{t}} = (I\otimes I - itH_{S}\otimes H_{A} - \frac{1}{2}t^{2}H_{S}^{2}\otimes H_{A}^{2})\ket{\Psi_{0}}
    \end{equation}
    When evaluating the effects of taking a quantum measurement, it's important to remember these following properties:
    \begin{enumerate}
        \item Quantum measurements can be described by a set of quantum operators \(\{M_{q}\}\)
        \item If the initial state of a quantum system is \(\ket{\psi}\), then the probability that result \(q\) occurs is: \(p(q) = \braket{\psi|M_{q}^{\dagger}M_{q}|\psi}\)
        \item the state of the system after the quantum measurement will become: \(\frac{M_{q}\ket{\psi}}{\sqrt{\braket{\psi|M_{q}^{\dagger}M_{q}|\psi}}}\)
    \end{enumerate}
    In order for us to properly take a weak measurement of the system, we want to take a specific projective measurement that only acts on the ancilla state. Such a measurement can be described by
    \begin{equation}
        E_{q} = I\otimes\ket{q}\bra{q}
    \end{equation}
    Projective measurements have the property that when applied to a composite system \(\ket{\psi}\): \(p(q) = \braket{\psi|E_{q}|\psi}\) and the post-measurement system becomes 
    \begin{equation}
        \frac{E_{q}\ket{\psi}}{\sqrt{p(m)}}
    \end{equation}
    Applying this measurement operator to our time evolved state, the post measurement state becomes:
    \begin{align}
        & \ket{\Psi_{q}} = \frac{E_{q}\ket{\Psi_{t}}}{\sqrt{\bra{\Psi_{t}}{E_{q}\ket{\Psi_{t}}}}} \\
        &= \frac{I\braket{\varphi|q} - itH_{S}\braket{q|H_{A}|\varphi} - \frac{1}{2}t^{2}H_{S}^{2}\braket{q|H_{A}^{2}|\varphi}}{\sqrt{\bra{\Psi_{t}}{E_{q}\ket{\Psi_{t}}}}} \ket{\psi}\otimes\ket{q}
    \end{align}
    What we can do now is set
    \begin{equation}
        M_{q} := I\braket{\varphi|q} - itH_{S}\braket{q|H_{A}|\varphi} - \frac{1}{2}t^{2}H_{S}^{2}\braket{q|H_{A}^{2}|\varphi}
    \end{equation}
    where \(M_{q}\) is called a Kraus operator.
    Returning back to deriving the weak measurement, we can describe our time evolved state with our Kraus operator below
    \begin{equation}
        \ket{\Psi_{q}} = \frac{E_{q}\ket{\Psi_{t}}}{\sqrt{\bra{\Psi_{t}}{E_{q}\ket{\Psi_{t}}}}} = \frac{M_{k}\ket{\psi}}{\sqrt{\bra{\psi}M_{k}^{\dagger}M_{k}\ket{\psi}}}\otimes\ket{\varphi}
    \end{equation}
    Note that the above postmeasurement state is a tensor product state, and therefore we may consider the state of the system alone, which is
    \begin{equation}
        \ket{\psi_{q}} = \frac{M_{k}\ket{\psi}}{\sqrt{\bra{\psi}M_{k}^{\dagger}M_{k}\ket{\psi}}}.
    \end{equation}
    By using the unitary evolution of the joint Hilbert space, we see that projective measurements can satisfy the properties of quantum measurements as we've previously established.

    We use a Krauss operator of the form
    \begin{equation}
        K_{\theta} = k \int_{-\infty}^{\infty} dX_{\theta} e^{-(X_{\theta}-\mu)^{2}/w^{2}}\ket{X_{\theta}}\bra{X_{\theta}},
    \end{equation}
    where $k$ is a normalization constant that will not appear in the postmeasurement state. We apply the resolution of identity,
    \begin{equation}
        \mathbb{I}= \int dx\ket{x}\bra{x}
    \end{equation}
    To represent the Kraus operator in the position basis:
    \begin{equation}
    \braket{x|K_\theta|x'}=k\int_{-\infty}^{\infty} dX_\theta e^{-(X_\theta-\mu)^2/w^2}\braket{x|X_\theta}\braket{X_\theta|x'}.
    \end{equation}
    From \ref{rotated_wavefunction} we can express this as,
    \begin{align}
        & \braket{x|K_\theta|x'}=\frac{k}{2\pi\hbar\sin\theta} \int_{-\infty}^\infty dX_\theta e^{-(X_\theta-\mu)^2/w^2}\exp\left(-\frac{i}{2\hbar}(x^2-x'^2)\cot(\theta)+\frac{i}{\hbar}(x-x')X_\theta \csc(\theta)\right) \\
        &= \frac{k}{2\pi\hbar\sin\theta}w\sqrt{\pi}\exp\left(\frac{1}{4\hbar^2} (x-x')\csc(\theta)[4 i \mu \hbar - 2 i (x+x')\hbar \cos(\theta)+w^2(x'-x)\csc(\theta)]\right)
    \end{align}

    Kraus operators are useful in scenarios such that if a coupled or joined system undergoes unitary evolution, we can use Kraus operators to describe how a specific subset of the system evolves in time.
    For example, if we take an initial system \(\ket{\psi}\otimes\ket{\varphi}\), it will evolve as follows
    \begin{equation}
        \ket{\psi}\otimes\ket{\varphi} \rightarrow \frac{M_{k}\ket{\psi}}{\sqrt{\bra{\psi}M_k^\dag M_k \ket{\psi}}}\otimes\ket{\varphi}
    \end{equation}
    If we now want to analyze just \(\ket{\psi}\), we can represent the initial state through density matrices.
    \begin{equation}
        \ket{\psi}\bra{\psi}\otimes\ket{\varphi}\bra{\varphi}\rightarrow \sum_{k,l}M_{k}\ket{\psi}\bra{\psi}M_{l}^{\dagger} \otimes \ket{\varphi_{k}}\bra{\varphi_{l}}
    \end{equation}
    By performing a partial trace with respect to \(\varphi\),
    \begin{align}
        &\ket{\psi}\bra{\psi}\rightarrow\sum_{k,l}M_{k}\ket{\psi}\bra{\psi}M_{l}^{\dagger}\braket{\varphi_{l}|\varphi_{k}} \\
        &= \sum_{k}M_{k}\ket{\psi}\bra{\psi}M_{k}^{\dagger}
    \end{align}
    Unlike the combined system which evolves under unitary transformation, a subsystem evolution can be described as
    \begin{equation}
        \rho\rightarrow\sum_{k}M_{k}\rho M_{k}^{\dagger}
    \end{equation}

\end{document}